\documentclass[twocolumn,aps,prl,superscriptaddress,reprint,longbibliography]{revtex4-1}
\usepackage[latin9]{inputenc}
\usepackage{amsmath,tikz,graphicx}
\usepackage{esint}
\usepackage{wasysym}
\usepackage[colorinlistoftodos]{todonotes}
\makeatletter
\usepackage{H1}

\renewcommand{\ket}[1]{|{{#1}}\rangle}

\makeatother

\begin{document}

\title{Comment on ``Entanglement growth in diffusive systems''}

\author{Tibor Rakovszky}

\affiliation{Department of Physics, Stanford University, Stanford, CA 94305, USA}

\affiliation{Department of Physics, T42, Technische Universit{\"a}t M{\"u}nchen, James-Franck-Stra{\ss}e 1, D-85748 Garching, Germany}

\author{Frank Pollmann}

\affiliation{Department of Physics, T42, Technische Universit{\"a}t M{\"u}nchen, James-Franck-Stra{\ss}e 1, D-85748 Garching, Germany}

\author{C.W.~von~Keyserlingk}

\affiliation{University of Birmingham, School of Physics \& Astronomy, B15 2TT,
UK}

\begin{abstract}

In a recent paper (Commun. Phys. 3, 100) \v{Z}nidari\v{c} studies the growth of higher R\'enyi entropies in diffusive systems and claims that they generically grow ballistically in time, except for spin-$1/2$ models in $d=1$ dimension. Here, we point out that the necessary conditions for sub-ballistic growth of R\'enyi entropies are in fact much more general, and apply to a large class of systems, including experimentally relevant ones in arbitrary dimension and with larger local Hilbert spaces. 

\end{abstract}

\maketitle

Recent works~\cite{RPvK19,Yichen} argued that R\'enyi entropies $S_\alpha$ with indices $\alpha > 1$ exhibit a sub-ballistic, $\propto \sqrt{t}$, growth in systems with diffusive transport.  A subsequent work, Ref. \onlinecite{Znidaric2020}, claims that such sub-ballistic growth occurs only in certain cases (in particular, $d=1$ dimensional systems with $q=2$ states per site) and is generically replaced by ballistic growth. Below, we argue that the diffusive entropy growth in fact pertains to a much wider class of systems. As we now detail, the conditions needed for diffusive (rather than ballistic) growth are incorrectly characterized in Ref. \onlinecite{Znidaric2020}, and are already apparent in the earlier discussions of Refs. \onlinecite{RPvK19,Yichen}. In particular, the example of Ref. \onlinecite{Znidaric2020} only avoids diffusive growth due to a particular symmetry structure, where only the spin on a single leg of a two-leg ladder is conserved.

\paragraph{1.} Ref. \onlinecite{Znidaric2020} considers U$(1)$-symmetric Floquet systems and claims that to have $S_{\alpha>1} \sim \sqrt{t}$ requires that all on-site diagonal operators (in some preferred basis) correspond to conserved quantities, with transport behavior that is diffusive (or slower). This is automatically satisfied in a system with $q=2$ states per site (e.g., a spin-$1/2$ system) with a single U$(1)$ symmetry (e.g., $\sum_j S_j^z$ being conserved), but it is not the case for $q>2$, leading to the claim that in such systems, $S_{\alpha>1} \sim t$, unless additional symmetries are present (e.g., if $\sum_j\left(S^z_j\right)^2$ is also separately conserved). We now present evidence that this statement is incorrect and that generically the conservation of $S^z$ is sufficient to induce $\sqrt{t}$ growth for arbitrary finite $q$. We also discuss which assumptions made in Ref.~\onlinecite{Znidaric2020} we expect to be responsible for this disagreement (see point $2$ below).

A direct refutation of the above claim is obtained by evaluating $S_2$ in a system with $q=3$. This is readily achieved in a random circuit model, extending the results of Ref.~\onlinecite{RPvK19} where the same was done for a $q=2$ chain. To be concrete, we consider a chain where the on-site Hilbert space resembles a Hubbard model in the infinite interaction limit (i.e., with double occupancies projected out): the three on-site states correspond to an empty site ($\ket{0}$), or  a site occupied by a spin-up/spin-down particle ($\ket{\uparrow}$,$\ket{\downarrow}$ respectively). We evolve the system with a brick-wall circuit of 2-site random unitaries which conserve the total number of particles but \emph{not} the spin. In this case,  Ref.~\onlinecite{Znidaric2020} would predict $S_{\alpha>1 }\sim t$  because $q=3$ and there is only a single U$(1)$ symmetry. On the contrary, calculating the \emph{annealed average} of $S_2$ numerically (see Ref. \onlinecite{RPvK19} for details) we find $S_2^{(a)}\propto\sqrt{t}$ for initial states that are superpositions of both empty and occupied sites (if all sites are occupied then the circuit precisely reduces to a $q=2$ random circuit \emph{without} symmetries, which has $\propto t$ growth~\cite{RvK17,Nahum17}; however, such initial states are finely-tuned).

\begin{figure}[h!]
	\includegraphics[width=0.65\columnwidth]{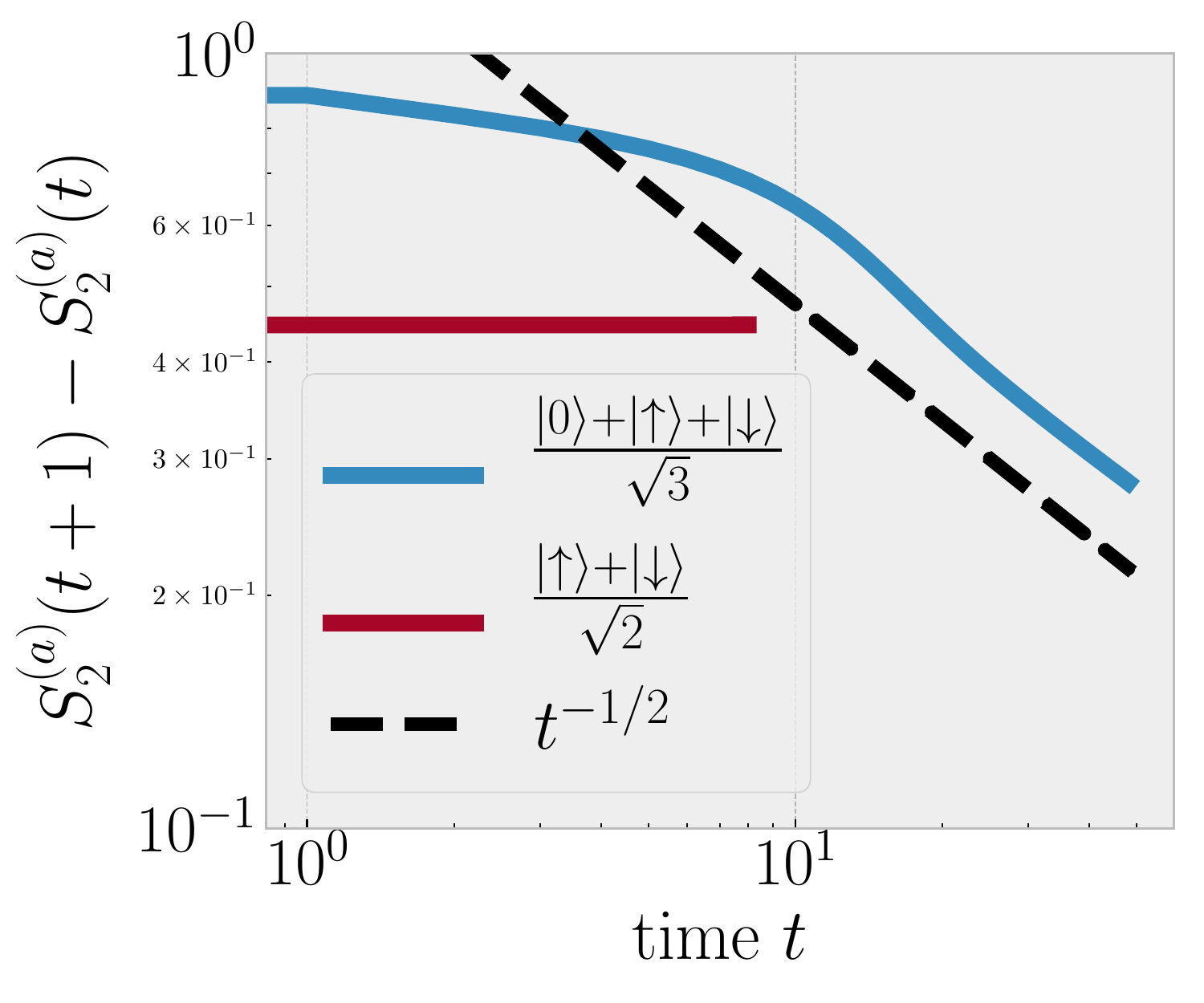}
	\caption{Time derivative of the half-chain annealed average $S_2$ in a random circuit with $q=3$ states per site and only a single U$(1)$ symmetry (see text). The red curve corresponds to an initial state which is a superposition of spin-up and spin-down particles only, with no empty sites; in this case the derivative is constant (ballistic). For a generic initial state that is the superposition of all three possible states (blue curve) the derivative decays as $1/\sqrt{t}$ (diffusive).}
	\label{fig:diffusive_growth}
\end{figure}

In fact, this result is expected: the proof of $\sqrt{t}$ growth (up to at most logarithmic corrections) originally derived for $q=2$ in Ref. \onlinecite{Yichen} extends straightforwardly to this model,and to many other $q>2$ systems, as pointed out recently in Ref. \onlinecite{YichenAddendum}. Rather than the particular value of $q$, the relevant condition for the proof is the existence of `empty' regions where no dynamics can occur due to the symmetry; let us call this the \emph{frozen region condition} (FRC)~\footnote{Nonetheless, the timescales necessary to observe the sub-ballistic growth might depend on $q$ and diverge as $q\to\infty$.}. For example, in our $q=3$ model, in an empty region, the state cannot evolve until some particles propagate into the region from the outside. 

To be more precise, we define the FRC in the following way. Consider a system of some finite size and make use of the symmetries to block-diagonalize the time-evolution operator. We say that the FRC is satisfied is there exist blocks containing only a single state for any system size. Entanglement growth should be at most $\propto \sqrt{t\log{t}}$ for any such system, by the arguments of Refs. \onlinecite{Yichen,YichenAddendum,RPvK19}.

The FRC is satisfied for a large class of systems, including many experimentally relevant ones. For example, it holds for any model where the local degree of freedom is a spin-$S$ variable and $\sum_j S_j^z$ is conserved: states that are fully polarized in the $z$ directions are frozen. As seen above, it also applies to systems of fermionic particles (or hard-core bosons), as long as their total number is conserved. On the other hand, this discussion highlights why certain systems \emph{do} have $S_{\alpha>1}\sim t$ despite their U$(1)$ symmetry: it can be the case that the symmetry only acts on some subset of the degrees of freedom, while others are unconstrained by it, such that no frozen regions can exist. This happens for example in a two-leg ladder if only the magnetization on one of the legs is conserved, as was shown in Refs. \onlinecite{RPvK19,Znidaric2020}. While such exceptions exist, they are in fact much less common than what would be implied by the conditions stated in Ref. \onlinecite{Znidaric2020}.

\paragraph{2.} Let us comment on the source of this disagreement. Ref. \onlinecite{Znidaric2020} provides a non-rigorous theoretical argument, aiming to connect the growth of $S_2$ to the decay of correlations of diagonal operators. The assumption made in this argument is that, while densities of explicitly conserved quantities (e.g. $S_j^z$) have power-law decaying correlations, most diagonal operators are insensitive to the symmetries and would decay exponentially. For example, under the assumptions of Ref. \onlinecite{Znidaric2020}, $\left(S_j^z\right)^2$ would have exponential correlations, unless $\sum_j \left(S_j^z\right)^2$ is explicitly conserved~\footnote{More precisely, it is better to consider the traceless version of the operator, $\tilde{S}_j^z \equiv 3 \left(S_j^z\right)^2 - 2$.}; this leads to the statement about $S_2 \sim t$. 

We believe that in fact the conservation of $\sum_j S_j^z$ is sufficient to cause power law decaying correlations (`hydrodynamic tails') in $\left(S_j^z\right)^2$. For example, $\left(S_j^z\right)^2$ can evolve into operators of the form $S_k^z S_l^z$ ($l \neq k$), i.e. the product of two conserved densities, which decay as $t^{-d}$ in a diffusive system. This is similar to the standard arguments that show the existence of long-time tails in the current operator itself~\cite{Pomeau1975}. For this reason, we expect that the class of operators with power-law decaying correlations is much larger than expected by Ref. \onlinecite{Znidaric2020}, which helps explain why $S_{\alpha>1} \sim \sqrt{t}$ is in fact much more prevalent. 

\paragraph{3.} Apart from the role of the size of the local Hilbert space $q$, Ref. \onlinecite{Znidaric2020} also raises the question about the validity of the $\sqrt{t}$ result in dimensions $d>1$. The numerical results of Ref. \onlinecite{Znidaric2020} are in fact consistent with the claim, made in Ref. \onlinecite{RPvK19}, that the sub-ballistic growth is present in any dimension. However, the way these results are presented could make this appear to be a finite-size effect. This is due to a choice of units: Ref. \onlinecite{Znidaric2020} measures time in units that depend on the overall system size (effectively rescaling $t \to t L^{d-1}$). In the more standard time units set by the microscopic couplings, the diffusive growth  sets in at a system-size independent timescale. 

\paragraph{4.} Let us make one final remark about the interpretation of these results that we believe might be confusing for readers of Ref.~\onlinecite{Znidaric2020}. There, it is claimed that one can think of $e^{S_2}$ as a measure of the number of degrees of freedom needed to describe the corresponding state. If this were so, the result $S_2 \sim \sqrt{t}$ would be rather powerful, indicating that systems obeying the FRC are much easier to simulate on a classical computer than other types of dynamics. However, this is not so. It was one of the important insights of Refs. \onlinecite{RPvK19,Yichen} that the long-time dynamics of $S_{\alpha>1}$ is dominated by the largest eigenvalue of the reduces density matrix. As such, knowing about these higher entropies tells us very little about the full complexity of the state; the lower R\'enyi entropies ($S_{\alpha \leq  1}$) might give a better characterisation of the information in a state~\cite{Norbert}, and indeed these appear to grow linearly in time even in diffusive systems~\cite{RPvK19,HyungwonHuse}.

\bibliographystyle{apsrev4-1}
\bibliography{comment.bbl}

\end{document}